\documentclass[sigconf]{acmart}
\AtBeginDocument{%
  }

\acmConference[ICSE '26]{2026 IEEE/ACM International Conference on Software Engineering}{April 12--18, 2026}{Rio de Janeiro, Brazil}
\acmYear{2026}

\acmDOI{}   
\acmISBN{}  

\usepackage{tabularx}
\usepackage[table]{xcolor}
\usepackage{booktabs}
\usepackage{caption}

\usepackage{subcaption}  

\usepackage{array}                        
\usepackage{pifont}

\usepackage{amssymb}
\usepackage{graphicx}
\usepackage{multirow} 
\usepackage{makecell}

\usepackage{titlesec}
\titlespacing{\section}{0pt}{*0.99}{*0.8}
\titlespacing{\subsection}{0pt}{*0.88}{*0.75}
\titlespacing{\subsubsection}{0pt}{*0.7}{*0.7}

\definecolor{green(pigment)}{rgb}{0.0, 0.65, 0.31}

\usepackage{tikz}

\usepackage{tikz}
\DeclareRobustCommand{\bcircled}[2][2.5ex]{%
  \tikz[baseline=(c.base)]\node[
    circle, fill=black, draw=black,
    minimum size=#1, inner sep=0pt,
    text=white, font=\bfseries
  ] (c) {#2};}

\usepackage{enumitem}
\newcounter{callout}

\begin{document}

\title{Software Vulnerability Management in the Era of Artificial Intelligence: An Industry Perspective}


\author{M. Mehdi Kholoosi}
\affiliation{%
  \institution{School of Computer Science and Information Technology}
  \institution{Adelaide University}
  \city{Adelaide}
  \country{Australia}
}
\email{mehdi.kholoosi@adelaide.edu.au}

\author{Triet Huynh Minh Le}
\affiliation{%
  \institution{School of Computer Science and Information Technology}
  \institution{Adelaide University}
  \city{Adelaide}
  \country{Australia}
}
\email{triet.h.le@adelaide.edu.au}

\author{M. Ali Babar}
\affiliation{%
 \institution{School of Computer Science and Information Technology}
 \institution{Adelaide University \&}
 \institution{Elevexai Systems}
  \city{Adelaide}
  \country{Australia}
}
\email{ali.babar@adelaide.edu.au}

\renewcommand{\shortauthors}{Kholoosi et al.}

\begin{abstract}
Artificial Intelligence (AI) has revolutionized software development, particularly by automating repetitive tasks and improving developer productivity. While these advancements are well-documented, the use of AI-powered tools for Software Vulnerability Management (SVM), such as vulnerability detection and repair, remains underexplored in industry settings.
To bridge this gap, our study aims to determine the extent of the adoption of AI-powered tools for SVM, identify barriers and facilitators to the use, and gather insights to help improve the tools to meet industry needs better. We conducted a survey study involving 60 practitioners from diverse industry sectors across 27 countries. The survey incorporates both quantitative and qualitative questions to analyze the adoption trends, assess tool strengths, identify practical challenges, and uncover opportunities for improvement. 
Our findings indicate that AI-powered tools are used throughout the SVM life cycle, with 69\% of users reporting satisfaction with their current use. Practitioners value these tools for their speed, coverage, and accessibility. However, concerns about false positives, missing context, and trust issues remain prevalent. We observe a socio-technical adoption pattern in which AI outputs are filtered through human oversight and organizational governance. To support safe and effective use of AI for SVM, we recommend improvements in explainability, contextual awareness, integration workflows, and validation practices. We assert that these findings can offer practical guidance for practitioners, tool developers, and researchers seeking to enhance secure software development through the use of AI.
\end{abstract}



\maketitle

\section{Introduction}
Software Vulnerabilities (SVs) are critical security issues that can result in cybercrime and substantial financial losses \cite{sanchez2019software}.
Beyond their immediate economic consequences, SVs can jeopardize data privacy and intellectual property, ultimately undermining an organization's overall security posture \cite{anwar2018understanding}.
The number and complexity of SVs have steadily increased, posing significant risks to software systems \cite{kholoosi-timing, le2022survey}. Despite these growing threats, many SVs remain unpatched for prolonged periods, with half taking over a year to resolve \cite{iannone2022secret,le2024latent}.
Tackling these challenges requires innovative automated techniques to alleviate the costly and time-consuming manual effort involved in managing SVs.
Software Vulnerability Management (SVM) involves a comprehensive set of processes aimed at improving the security of software systems through the systematic detection, assessment, repair, and disclosure of SVs \cite{dowd2006art, le2022survey}.
These phases are collectively referred to as the SVM life cycle \cite{shahzad2019large,le2022towards}, and we use this definition throughout this paper.

Recent studies have increasingly proposed Artificial Intelligence (AI)-powered tools to support the SVM life cycle. Among these, Deep Learning (DL) has emerged as a prominent trend, showing promising results for Software Engineering (SE) \cite{le2020deep}, including various SVM tasks.
For instance, DL models have shown effectiveness in detecting and assessing SVs in different application domains~\cite{tang2023deep,cao2023learning,deepcva,nguyen2024automated}. 
Unlike traditional static analysis tools that rely on predefined rules, these models learn vulnerability patterns during training, enhancing detection accuracy \cite{tang2023deep, cao2023learning}.
DL-based Automated Program Repair (APR) has also shown notable progress. These APR techniques leverage neural networks to identify bug-fixing patterns, achieving remarkable performance in various scenarios \cite{zhang2023survey, sintaha2023katana, tian2023best}. The emergence of pre-trained DL models (e.g., CodeBERT \cite{feng2020codebert}) has further boosted SVM research efforts, yielding superior results compared to traditional approaches. Such models have demonstrated high vulnerability repair accuracy (32.94–44.96\%) \cite{zhang2023pre}, and domain adaptation techniques have improved APR effectiveness by up to 48.78\% \cite{zirak2024improving}. Moreover, Vision Transformer-inspired methods have achieved better performance in automated vulnerability repair, outperforming baselines by 2.68–32.33\% \cite{fu2024vision}. Large Language Models (LLMs) have also demonstrated potential in reproducing bugs \cite{kang2024evaluating} and repairing hardware security bugs~\cite{ahmad2024hardware} as well as misconfigurations~\cite{ye2025llmsecconfig}.

However, despite promising results in controlled settings commonly evaluated in academic studies, AI-powered tools, particularly those based on DL, often struggle in real-world scenarios.
Studies have reported performance drops of up to 95 percentage points when evaluating these models on realistic datasets \cite{chakraborty2024revisiting}. Challenges such as overfitting, dataset imbalance, and suboptimal model choices continue to hinder the practical applicability of DL-based vulnerability management solutions \cite{croft2023data, chakraborty2021deep,le2024automatic,le2024mitigating,11185080}.
Additionally, a wide range of AI-powered security products are already available on the market \cite{AI_Security_Companies}, yet it is still unclear to what extent software security practitioners actively use these tools.

Given the potential and challenges of AI-powered tools, as well as the uncertainty surrounding their industry adoption, it is crucial to understand how they are used in real-world SVM practices.
The rapid advancement of AI technologies and their growing adoption in software development workflows \cite{Sonatype_AI_Software_Development_2023} further underscore the urgency of this investigation. 

To address this gap, we conducted an empirical study using an online questionnaire in March-May 2025, targeting experienced software security professionals (N=60). Our study examines current practices across the SVM life cycle, including the AI-powered tools in use, perceived strengths, and challenges practitioners face. We also explore how they integrate AI outputs into their workflows and the strategies they employ to verify them.

To guide our investigation, we answer two research questions:
\begin{itemize}
    \item{\textbf{RQ1:} How are AI-powered tools currently being used by software security practitioners across the SVM life cycle?}

    \item{\textbf{RQ2:} What are the perceived strengths and limitations of AI-powered tools for SVM in industry settings?}
\end{itemize}

To the best of our knowledge, no prior study has systematically investigated how AI-powered tools are adopted and applied across the full SVM life cycle from a practitioner perspective. Existing studies often focus on specific tools, tasks, or development contexts, whereas our work offers the first holistic, industry-wide view grounded in real-world practices (see Section~\ref{sec:Related_Work}). Our findings provide actionable insights into how AI is shaping security practices, as well as what must evolve to enhance its utility in the future. All study materials are available in the supplementary package \cite{figshare}.

\section{Related Work}
\label{sec:Related_Work}
\color{black}
This section reviews prior work at two levels. Section~\ref{Sec:AI4SE-beyondSVM} summarizes developer-perception studies on AI for SE. Section~\ref{Sec:Developer-centered} synthesizes developer-centered studies on AI use in SVM. \color{black}

\color{black}
\subsection{\textbf{Developer perceptions in AI for SE}}
\label{Sec:AI4SE-beyondSVM}

Most prior developer-perception studies in AI for SE have focused on code generation and AI assistants.
Vaithilingam et al. \cite{vaithilingam2022expectation} found that GitHub Copilot \cite{GitHub-Copilot} did not necessarily improve programming task completion time or success rate, yet many participants preferred it for useful starting points while struggling to understand and debug generated code. 
Ziegler et al. \cite{ziegler2022productivity} reported that the rate at which developers accepted the assistant's code suggestions was the best predictor of developers' self-reported productivity. 
Liang et al. \cite{liang2024large} surveyed 410 developers and found that programmers primarily use AI assistants for practical efficiency gains like reducing key-strokes, finishing programming tasks quickly, and recalling syntax, while showing less interest in using them to brainstorm potential solutions; key barriers were assistants failing to generate code that meets functional or non-functional requirements and difficulties controlling the tool to generate desired output.
Sergeyuk et al. \cite{sergeyuk2025using} observed that 84.2\% of programmers used AI assistants at least occasionally, with use concentrated on implementing new features and on generating or summarizing code; they also found that developers most often delegated writing tests and natural-language artifacts to assistants, while reported barriers included inaccurate output, lack of trust and desire for control, and limited project-context understanding by the assistant.
Collectively, these studies found that in code generation tasks, AI assistants provided scaffolding and were associated with higher perceived productivity, but remained constrained by trust, accuracy, control issues, and limited project context \cite{vaithilingam2022expectation,ziegler2022productivity,liang2024large,sergeyuk2025using}.
\color{black}

\color{black}\subsection{\textbf{Developer-centered studies on AI for SVM}}\color{black}
\label{Sec:Developer-centered}
Several developer-centered studies (i.e., a subset of empirical studies with developers that adopt a specific focus) \cite{fu2024aibughunter, liu2024exploring, klemmer2024using, steenhoek2025closing} have examined the intersection of AI and SVM. 
Table \ref{tab:related_work_comparison} provides a comparative overview of the existing developer-centered studies on AI use in SVM, highlighting their tool focus, task coverage, scope, and methodological limitations relative to our investigation.

\begin{table*}[ht]
\centering
\caption{Positioning our study in the literature: a comparison of developer-centered studies on AI use in SVM.
\newline \textit{SVM task abbreviations:} D = Detection, A = Assessment, R = Repair, Di = Disclosure}
\label{tab:related_work_comparison}
\rowcolors{2}{gray!20}{white}
\begin{tabularx}{\textwidth}{|>{\centering\arraybackslash}m{1.4cm}|
                                  >{\centering\arraybackslash}m{2.6cm}|
                                  >{\centering\arraybackslash}m{0.5cm}|
                                  >{\centering\arraybackslash}m{0.5cm}|
                                  >{\centering\arraybackslash}m{0.5cm}|
                                  >{\centering\arraybackslash}m{0.5cm}|
                                  >{\centering\arraybackslash}m{4.5cm}|
                                  >{\centering\arraybackslash}m{2.9cm}|
                                  >{\centering\arraybackslash}X|}
\hline
\rowcolor{gray!50}
\centering\arraybackslash\textbf{Study} & 
\centering\arraybackslash\textbf{AI Tool(s) Included} & 
\multicolumn{4}{c|}{\centering\arraybackslash\textbf{SVM Task Focus}} & 
\centering\arraybackslash\textbf{Scope of User Study} & 
\centering\arraybackslash\textbf{AI vs. Non-AI Tools Usage Frequency} & 
\centering\arraybackslash\textbf{\#N} \\
\rowcolor{gray!50}
& & \centering\arraybackslash\textbf{D} & \centering\arraybackslash\textbf{A} & \centering\arraybackslash\textbf{R} & \centering\arraybackslash\textbf{Di} & & \textbf{for SVM} & \\
\hline
\textbf{Fu et al.} \cite{fu2024aibughunter} & Only AIBugHunter & \checkmark & \checkmark & \checkmark & -- & Preliminary evaluation of AIBugHunter tool & No & N=27 \\
\textbf{Liu et al.} \cite{liu2024exploring} & Only ChatGPT & -- & -- & -- & \checkmark & Preliminary evaluation of ChatGPT & No & N=20 \\
\textbf{Klemmer et al.} \cite{klemmer2024using} & Only LLM-powered tools & \checkmark & -- & -- & -- & Not specifically centered on SVM workflows & No & N=27 \\
\textbf{Steenhoek et al.} \cite{steenhoek2025closing} & Only DEEPVULGUARD & \checkmark & -- & \checkmark & -- & Confined to a single organization (Microsoft) & No & N=17 \\
\hline
\rowcolor{cyan!15}
\textbf{Our Study} & Every shape or form of AI-powered tools & \checkmark & \checkmark & \checkmark & \checkmark & Industry-wide across the full SVM life cycle & Yes & N=60 \\
\hline
\end{tabularx}
\end{table*}

\color{black}
\textit{AI Tools and SVM Tasks.}
Existing work concentrates on individual tools or narrow task domains. Liu et al. \cite{liu2024exploring} conducted a comprehensive evaluation of ChatGPT’s performance across the complete SVM life cycle, benchmarking it against 11 state-of-the-art approaches. In addition to automated evaluation, they carried out a pilot user study, but it focused on the single task of vulnerability report summarization. Steenhoek et al. \cite{steenhoek2025closing} developed and assessed DEEPVULGUARD (a VS Code extension built on state-of-the-art models) for detection and repair, emphasizing capability within one implementation. Fu et al. \cite{fu2024aibughunter} presented AIBugHunter (a Machine Learning (ML)-based SV detection and repair plugin for C/C++ in VS Code); again, the evaluation was tied to a single IDE-integrated tool within a constrained language. Klemmer et al. \cite{klemmer2024using} provided a broader look at general-purpose AI assistants (e.g., ChatGPT, Copilot) in secure development, but the focus is on general use rather than systematic SVM workflows or adoption trends.

\textit{Study Scope and Samples.}
These studies rely on controlled or local settings with limited participant pools. Steenhoek et al. \cite{steenhoek2025closing} reported results from 17 professional developers within one organization (Microsoft). Fu et al. \cite{fu2024aibughunter} included a small lab study (6 developers under timed tasks) plus a survey of 21 practitioners. Klemmer et al. \cite{klemmer2024using} interviewed 27 professionals and analyzed Reddit posts, providing qualitative insight without systematic SVM life cycle coverage. Liu et al. \cite{liu2024exploring} supplemented their benchmarking with a pilot user study on bug report summarization task, recruiting 20 participants (5 industry experts and 15 open-source maintainers).

\textit{Gaps and Positioning.}
\color{black}
None of these studies quantified the relative usage frequency of AI versus non-AI tools for SVM tasks. Across these studies, evidence is anchored to single tools or narrowly scoped tasks, typically assessed in lab-style benchmarks or single-organization case studies, leading to fragmented coverage of detection, assessment, repair, and disclosure, and no mapping of industry-wide adoption, governance, or organizational integration for AI in SVM. \color{black} In contrast, our study offers a broader, practitioner-focused perspective on AI use in SVM. Note that our investigation is not limited to DL- or LLM-based tools; instead, we consider all types of AI tools that involve data-driven decision making. With the largest pool of participants among related studies (N=60) spanning 27 countries and representing a wide range of roles and levels of experience, we can provide a comprehensive view of real-world AI practices for SVM. Specifically, by surveying this diverse group, we aim to capture current usage patterns, understand perceived strengths, and identify adoption barriers across the entire SVM life cycle. This broader lens helps uncover systematic challenges and practical opportunities that may be overlooked in tool- or task-specific studies, providing actionable insights for practitioners, researchers, and tool developers working in the field of AI for SVM.

\section{Methodology}

\subsection{Recruitment and Selection Criteria}
\label{sec:Recruitment}
We specifically recruited software security practitioners to capture a diverse range of security experiences across the SVM life cycle. To ensure relevance and expertise, participants were required to meet two eligibility criteria: (1)  professional experience with tools or methods used in the SVM life cycle (i.e., identifying, assessing, mitigating, or reporting SVs in source code), and (2) a minimum of one year of professional experience in their field.

\subsubsection{\textbf{Recruitment Channel.}}
We recruited participants through crowdsourcing platforms, as recommended by prior research, which highlights their effectiveness in accessing a large pool of participants for programming-related studies \cite{liang2024large, tahaei2022recruiting}. We recruited participants via Freelancer.com, a crowdsourcing platform. Freelancer has been previously recommended as an effective resource for recruiting participants in security-centric research \cite{kaur2022recruit} and has been used in several usable security studies involving developers \cite{naiakshina2020conducting, geierhaas2022let}. We posted our project nine times on the platform, with each posting active for one week. This approach allowed us to gradually collect responses, monitor data quality, and manage workload during the validation process (Section~\ref{sec:Validation_Responses}). It also ensured that we could stop recruitment once thematic saturation was reached (Section~\ref{sec:Qualitative Analysis}). Every call attracted a large number of freelancers expressing interest in participating. In this study, we 
opted not to implement a 
screening questionnaire. Recent research has raised concerns about the misuse of AI chatbots in survey settings, as such tools can be employed to automatically generate responses, potentially undermining the integrity and validity of crowdsourced data collection \cite{wang2023safeguarding, rothschild2024opportunities}. To mitigate this risk, we implemented a strict manual vetting protocol during participant selection (Section~\ref{Sec:Selection_Protocol}). Furthermore, we conducted a rigorous post-submission validation phase to assess the quality and authenticity of responses (Section~\ref{sec:Validation_Responses}).

\subsubsection{\textbf{Selection Protocol.}}
\label{Sec:Selection_Protocol}
Selection was limited to freelancers who held two 
verification badges on their Freelancer.com profiles: the Verified Freelancer badge and the Preferred Freelancer badge. The Verified Freelancer badge (indicated by a blue checkmark) is granted 
after the freelancer completes 
an identity verification process, including a video interview conducted by Freelancer staff. This ensures the authenticity of the account holder. The Preferred Freelancer badge (represented by a gold origami bird icon) is awarded to top-tier freelancers based on 
performance, professionalism, and subject-matter expertise. Entry into the Preferred Freelancer Program is 
competitive and involves internal vetting based on prior project success, client satisfaction, and demonstrated skills. 

In addition to badge verification, we manually reviewed each candidate’s profile, including their self-described expertise, client reviews, and portfolio of completed projects. To further assess practical experience, we initiated one-on-one conversations with shortlisted candidates via Freelancer’s real-time chat interface. During these conversations, we asked about their hands-on experience with tools or methods used for SVM tasks in source code. Only candidates who demonstrated relevant expertise were invited to participate and provided with the survey instructions and participation link. During enrollment, participants were informed that their responses would undergo a post-submission validation phase. Specifically, they were notified that we would assess the quality of their answers and review them for AI-generated content. We provide further details about this validation process in Section \ref{sec:Validation_Responses}. 

Across our nine project postings, we received 117 proposals from interested freelancers. In parallel, we used the platform’s advanced search feature to proactively identify and reach out to 436 potential candidates using relevant keywords such as “vulnerabilit*”. After applying our selection protocol to both recruitment approaches (i.e., proposal-based and advanced search), we ultimately invited 94 qualified freelancers to participate in the study. 

\begin{table*}[ht]
\centering
\caption{Overview of survey structure and question types by block. The complete survey is in the supplementary materials \cite{figshare}.}
\label{tab:survey_structure}
\begin{tabularx}{\linewidth}{|>{\raggedright\arraybackslash}m{2.8cm}
                        |>{\raggedright\arraybackslash}m{9.4cm}
                        |>{\raggedright\arraybackslash}X|}
\hline
\rowcolor{gray!40}
\multicolumn{1}{|c|}{\textbf{Block}} & 
\multicolumn{1}{c|}{\textbf{Sample Question}} & 
\multicolumn{1}{c|}{\textbf{Number and Type of Questions}} \\
\hline
Demographics & What is your current role in the industry? & 7 multiple-choice, 1 open-ended \\ \hline
SVM Experience & Which specific vulnerability management task(s) have you worked on? & 1 multiple-choice, 2 Likert-scale \\ \hline
AI Practices (\textbf{RQ1}) & Which AI-powered tools have you used for each of the SVM tasks? & 5 multiple-choice, 3 open-ended \\ \hline
Perceived Strengths \& Limitations (\textbf{RQ2}) & Based on your experience, what are the key strengths of AI-powered tools in managing software vulnerabilities? & 1 multiple-choice, 2 open-ended \\ \hline
\end{tabularx}
\end{table*}

\subsection{Survey Design}
\label{sec:Survey_Design}
To address our research questions, we developed 
a survey using Qualtrics \cite{qualtrics}. Our survey design was informed by 
guidelines for conducting survey research in software engineering \cite{linaaker2015guidelines}. The survey was developed through an iterative process and piloted in two rounds. The first round involved two experienced software security researchers (each with over three years of SVM research experience), who helped us refine the wording and improve the coverage of the questions. In the second round, we piloted the revised survey with four industry-based software security practitioners, each with a minimum of two years of professional experience in security-focused roles, to gather practical feedback and estimate survey duration. The survey was updated between each round, and responses from pilot participants were excluded from the final dataset. Prior to public release, the study received ethics approval from our institution’s Human Research Ethics Committee.

Following the piloting, 
the final version of the survey was structured into multiple blocks. Below, we describe the design of each block and how participants progressed through it. Each block appeared on at least one separate page, and backward navigation was disabled to preserve response integrity. \color{black} We adopted a forward-only design to preserve first-pass answers and avoid retrospective edits across blocks. After completion, participants received a thank-you email with a direct contact channel; no one requested changes, suggesting this design did not raise issues. 
An overview of the survey structure and question types by topic is provided in Table \ref{tab:survey_structure}. 

\subsubsection{\textbf{Welcome Page.}}
The first page provided participants with an overview of the study, data handling policies, and informed consent. Participants were advised that their participation was voluntary and that they could withdraw at any point.

\subsubsection{\textbf{Demographics.}}
We collected demographic data, including age, gender, education level, country of residence, current industry sector, professional role, years of experience, and organization size. 
All demographic questions were mandatory, as this data allows us to understand which segments of the security practitioner population are represented in our dataset.

\subsubsection{\textbf{SVM Experience.}}
This block gathered information on participants’ background related to SVM. To assess prior experience, we used a 5-point Likert scale (from “None at all” to “A great deal”) to measure involvement in each of the four core SVM tasks (i.e., SV detection, assessment, repair, and disclosure) \cite{dowd2006art, le2022survey}. We also asked participants to report how many hours per week they typically spend on vulnerability-related tasks (multiple choice). 

Next, we used a 9-point Likert scale (from “Never” to “Always”) to assess the frequency with which participants use existing approaches to manage SVs. Although this study primarily focuses on AI-powered tools, we included other commonly used approaches to understand how AI fits within the broader SVM landscape and to enable comparisons in adoption patterns. Based on prior research, we identified four main approaches that are commonly used individually or in combination for managing SVs in source code:

\begin{itemize} [itemsep=0pt, parsep=0pt, topsep=4pt, partopsep=0pt, after=\vspace{-0.09\baselineskip}]
   
    \item \textit{AI-powered tools:} These tools leverage ML and DL (including LLMs) models to automate management of SVs \cite{shiri2024systematic,fu2024aibughunter,steenhoek2025closing}.

    \item \textit{Application Security Testing (AST) tools:} AST tools assess code or behavior to manage SVs across the development pipeline. They include static, dynamic, and interactive analysis tools, with varying degrees of reliance on predefined rules versus behavioral or heuristic methods \cite{hanif2021rise,li2024static}.
  
    \item \textit{Manual analysis and review:} This involves direct inspection of code by practitioners to identify potential SVs \cite{hanif2021rise,allodi2020measuring}.  
  
    \item \textit{Outsourced services:} Some organizations delegate SVM tasks to external vendors or consultants \cite{alomar2020you,shahzad2012large,shahzad2019large}.
    
\end{itemize}

To minimize potential bias or misinterpretation due to unfamiliarity with these approaches, we provided concise definitions through a tooltip (hint icon) next to the question. These descriptions outlined the key characteristics of each SVM approach, ensuring participants could make informed selections.

\subsubsection{\textbf{AI Practices.}}
\label{sec:AI Practices}
This block focused on practitioners’ 
use of AI-powered tools across the SVM life cycle (RQ1). Participants were asked to name the 
tools they used for each task. They also reported how these tools influenced their decision-making, 
and the methods they used to integrate AI-generated outputs into their workflows. 

During the piloting phase, we observed LLMs were the most frequently mentioned type of AI-powered tool practitioners used for SVM tasks. This observation aligns with recent research highlighting the widespread popularity of LLMs among software practitioners for various development tasks~\cite{chen2025empirical}. Consequently, we added a set of questions to capture more nuanced insights into LLM use. Participants were asked to report how frequently they used LLMs for SVM tasks, which types of LLMs they primarily relied on, and what prompt engineering techniques they typically employed to enhance the effectiveness of these tools. Additionally, we inquired about the measures participants take to verify the reliability of LLM-generated outputs, given the risks associated with relying on AI-generated suggestions in security-critical contexts \cite{mousavi2024investigation}.

\subsubsection{\textbf{Perceived Strengths \& Limitations.}}  
This block focused on eliciting participants’ reflections on the benefits and challenges of using AI-powered tools for SVM, 
addressing RQ2. Participants were asked two 
questions: one inviting them to describe the key strengths of such tools based on their 
experience, and another exploring the challenges they had encountered. In addition, participants rated their overall satisfaction with AI-powered tools 
in SVM tasks.

\subsection{Validation of Responses}
\label{sec:Validation_Responses}

Of the 94 
freelancers invited, 67 completed the survey, producing a 71\% response rate, which is considered high for online surveys and indicative of the effectiveness of our recruitment strategy \cite{wu2022response}. 

To ensure the authenticity and quality of the collected data, we conducted a thorough post-submission validation process in which all responses were carefully reviewed. As part of this process, the recorded completion time served as an initial indicator of potential issues. During the pilot phase, we observed that the survey typically took around 35 minutes to complete. While all 67 responses were evaluated, any submission completed in under 30 minutes was flagged for closer scrutiny. We identified three such responses, each completed in approximately 15 minutes, that included unusually long, well-structured paragraphs in multiple answers to open-ended questions. These texts were written in flawless English, with perfect grammar and no spelling errors, raising strong suspicions of AI-generated content. Additional indicators of low-quality responses included repetitive Likert scale selections and generic answers to open-ended questions. For instance, some participants provided textbook-style definitions of vulnerability analysis methods rather than describing their own practical experiences. Four responses exhibited these patterns. The first author conducted the initial review of all 67 responses. Submissions flagged during this review were then independently assessed by the second author. Final exclusion decisions were made jointly, based on a thorough evaluation of response quality and consistency. As a result, seven responses were excluded from the final dataset.

Following this validation process, 60 high-quality responses were accepted into the final dataset, with 27 from the open proposal-based calls and 33 from the advanced search efforts. We determined that further data collection was unnecessary, as thematic saturation had been reached during our qualitative analysis (see Section~\ref{sec:Qualitative Analysis}). This rigorous, multi-stage recruitment and vetting process ensured that only qualified practitioners with demonstrable experience in SVM-related tasks were included in the study. Each participant was compensated \$50 AUD for their time via the Freelancer platform.

\subsection{Survey Analysis}
\label{sec:Survey_Analysis}

\subsubsection{\textbf{Quantitative Analysis.}}
\label{sec:Quantitative Analysis}
We analyzed responses to closed-ended questions following established guidelines for survey analysis in software engineering research~\cite{kitchenham2008personal}. For multiple-choice questions, we primarily reported descriptive statistics, including frequency counts and percentages indicating how often each response was selected. For Likert-scale items, we computed frequency distributions to capture patterns and identify trends in respondents' practices and perceptions. To assess differences in reported usage of SVM approaches (measured using a nine-point ordinal scale), we used the Friedman test~\cite{friedman1937use}, which is suitable for comparing ranked responses from the same participants, especially when the data do not follow a normal distribution. When significant differences were found, we conducted Wilcoxon signed-rank tests with Holm correction~\cite{holm1979simple} for post-hoc comparisons. We also calculated Kendall’s W~\cite{kendall1939problem} to assess the level of agreement in participant rankings.

\begin{figure*}[h]
  \centering
  \includegraphics[width=0.98\linewidth]{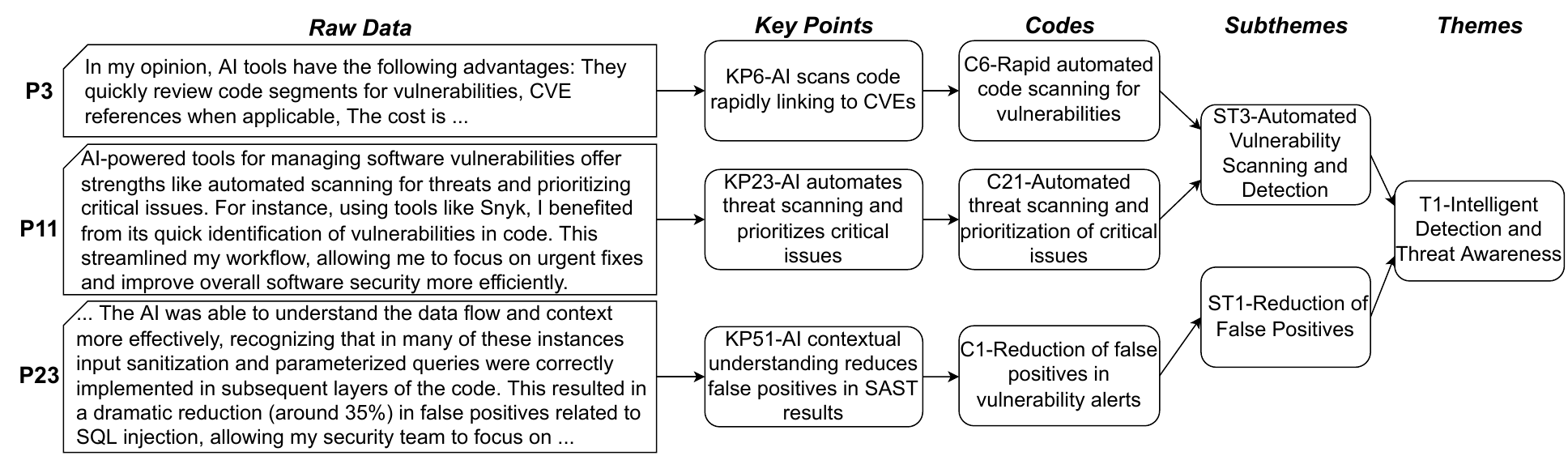}
  \caption{Illustration of our coding process for the question "What are the key strengths of AI-powered tools in SVM?".}
  \label{fig:Illustration}
\end{figure*}

\subsubsection{\textbf{Qualitative Analysis.}}
\label{sec:Qualitative Analysis}
For the analysis of open-ended survey questions, we followed Braun and Clarke’s six-step thematic analysis approach~\cite{braun2006using}, using an inductive (i.e., data-driven) strategy to generate a rich and nuanced account of the responses. \color{black}We selected this method because it is widely used in similar contexts \cite{klemmer2024using,van2020schrodinger,asthana2023case, tahaei2021security}. \color{black} Each open-ended question was analyzed independently, meaning each had its own distinct set of codes, subthemes, and themes. \color{black}{We adopted a pragmatic stance and conducted a \emph{semantic} thematic analysis focused on explicit participant statements rather than latent meanings.}\color{black}

We began by familiarizing ourselves with the responses for each question (Step 1). To initiate coding, we randomly selected six survey responses (10\% of the dataset), similar to established practices in related works~\cite{iwaya2023privacy,Kholoosi-chatgpt,jacobsen2025chatbots}. For each open-ended question, the first and second authors (with 5 and 10 years of experience in software security and AI-powered SV research, respectively) independently reviewed the responses and generated initial codebooks by first extracting key points and then assigning preliminary codes to them (Step 2). These initial codebooks were then compared, merged, and refined through discussion, aligning overlapping codes and resolving discrepancies. Next, the first author continued coding the remaining responses for each question using the consolidated codebooks, while documenting any new codes as they \color{black}were inductively developed\color{black}. Codes were then grouped into meaningful subthemes (Step 3), which were iteratively refined into broader themes for each question (Step 4). These themes were defined and named clearly to reflect their underlying concepts (Step 5), and illustrative examples were selected to support our reporting (Step 6). 
\color{black}To support reflexivity, we held \color{black}weekly meetings with the research team \color{black} (including the third author) \color{black} during which evolving codes, subthemes, and themes were discussed and refined to ensure clarity and consistency.

We reached thematic saturation across all open-ended questions after analyzing 46 responses, meaning that no new subthemes were identified 
after this point. Some questions reached saturation earlier in the process. On average, we assigned approximately 49 codes per open-ended question, reflecting the depth and granularity of our qualitative analysis. We did not calculate an inter-rater reliability (IRR) measure, consistent with established qualitative research guidelines \cite{mcdonald2019reliability,braun2022starting}. Our coding approach involved iterative discussion and refinement, emphasizing consensus and collaborative interpretation. Thus, the themes \color{black}were developed \color{black} through a negotiated agreement process, making IRR less relevant, as it typically assesses consistency of independent coding rather than interpretive agreement among coders. This approach is consistent with prior qualitative studies that similarly emphasize interpretive agreement over statistical coder alignment \cite{liang2024large,iwaya2023privacy,klemmer2024using}. 
Figure \ref{fig:Illustration} displays a sample from our coding process.

\section{Results}
\label{sec:Results}
In this section, we report the results of our investigation. For closed-ended questions, we present descriptive statistics and, where applicable, statistical tests to highlight key patterns in participant responses. For open-ended questions, we summarize our qualitative analysis through themes and subthemes derived from the data. To indicate the \color{black}prevalence \color{black} of each theme or subtheme, we report relative reference counts (\#Refs), which reflect the total number of keypoint references coded to that theme or subtheme across responses. We then discuss each theme in detail, focusing on the most \color{black} prevalent \color{black}subthemes (i.e., those with the highest reference counts) and supporting them with direct participant quotations. \color{black} Although theme labels are high-level and may not themselves signal SVM specificity, we constructed them from SVM-specific mechanisms evident at the subtheme level; the majority of subthemes are grounded in concrete SVM activities. Throughout this section, we explicitly flag SVM-specific subthemes as we present them. For readability, themes for open-ended questions are prefixed with a black circled marker (e.g., \bcircled{T1}, \bcircled{T2}). \color{black} The full set of codes, subthemes, and themes for each question is provided in the supplementary materials \cite{figshare}.

\subsection{Our Participants}
\label{sec:Participants}

\subsubsection{\textbf{Participant Demographics.}}
\label{sec:Participant_Demographics}
Respondents provided 
four socio-demographic factors: age, gender, location, and highest level of education. Most participants were between the ages of 25–34 (45\%) and 18–24 (\color{black}27\color{black}\%), suggesting 
the sample primarily comprised early to mid-career professionals. The gender distribution was overwhelmingly male (\color{black}97\color{black}\%). Participants were geographically distributed across six continents, with the largest groups from Asia (\color{black}32\color{black}\%), North America (\color{black}23\color{black}\%), and Europe (20\%). In terms of education, the majority held either a bachelor’s degree (\color{black}48\color{black}\%) or a master’s degree (\color{black}37\color{black}\%), indicating a highly educated practitioner base.

Respondents 
reported four employment-related details: professional role, years of experience, industry sector, and organization size. Approximately 70\% of participants 
holding security-focused roles, with the most common being Penetration Tester (\color{black}13\color{black}\%), Cybersecurity Consultant (\color{black}11\color{black}\%), Security Researcher (\color{black}11\color{black}\%), and Security Analyst (\color{black}11\color{black}\%). Fewer participants 
held general software or infrastructure roles, such as Software Architect (\color{black}8\color{black}\%) or Software Engineer (\color{black}6\color{black}\%). Regarding professional experience, the most common bracket was 2–5 years (\color{black}37\color{black}\%), with \color{black}43\color{black}\% reporting more than five years of experience. Based on ordinal midpoints, the average tenure was 
6.5 years, indicating a mid-career cohort. Overall, the sample reflects a practitioner base with substantial hands-on expertise in software security. In terms of industry affiliation, most participants reported working in cybersecurity services (\color{black}57\color{black}\%) or information technology (\color{black}28\color{black}\%), with smaller representation from finance, education, government, and defense. Organization sizes varied: \color{black}27\color{black}\% of participants reported working in companies with 1–10 employees, another \color{black}27\color{black}\% in organizations with more than 100 employees, and the remainder distributed across mid-range categories (11–100 employees).

\subsubsection{\textbf{SVM Experience.}}

We asked participants how much time they dedicate to SVM tasks each week. A majority (60\%) reported spending more than 10 hours weekly, with over one-third (36.6\%) dedicating more than 20 hours. These responses indicate that most participants are regularly engaged in SVM activities.

\begin{figure}[htbp]
  \centering
  \includegraphics[width=0.9\linewidth]{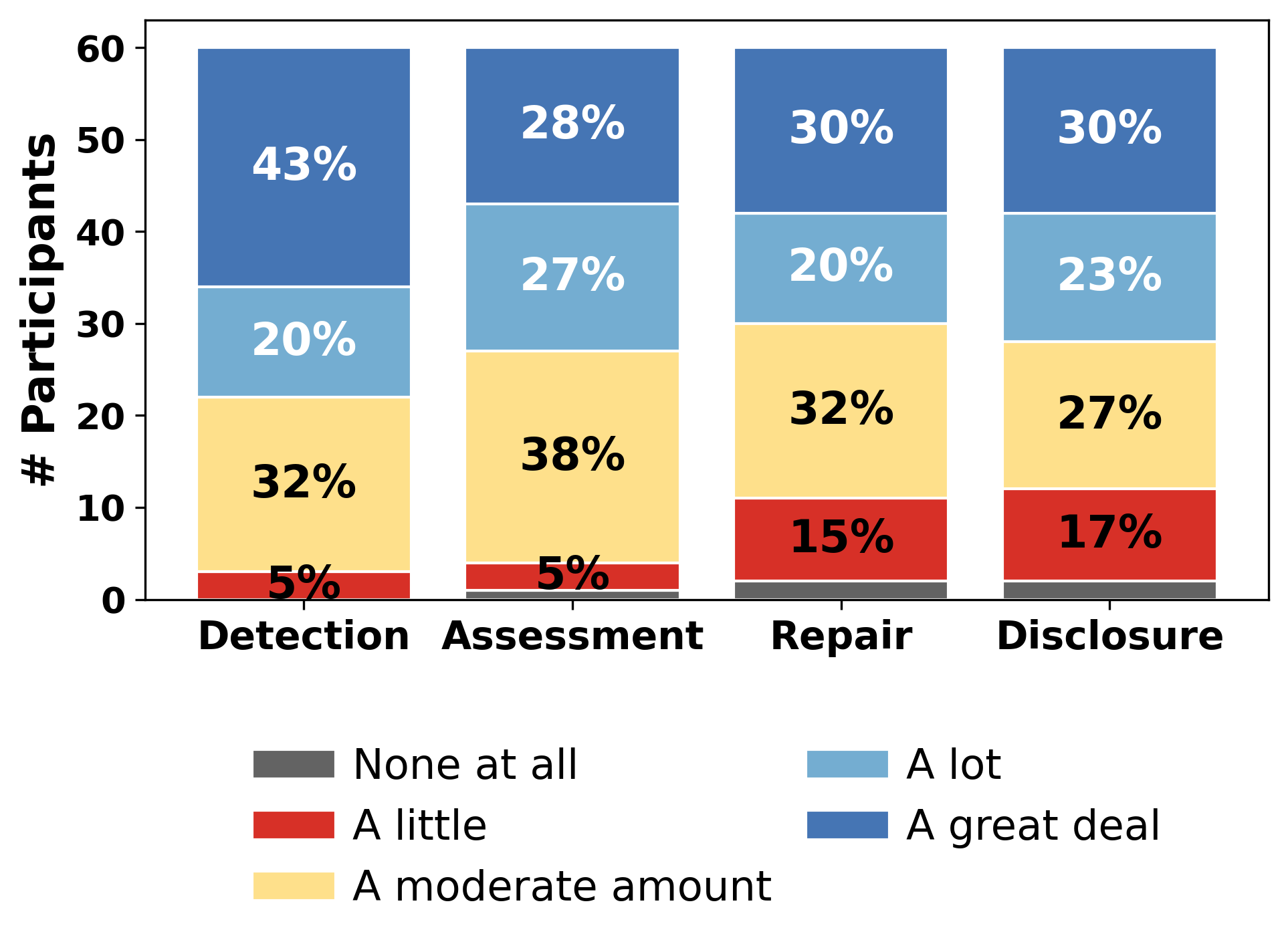}
  \caption{Participants’ involvement in key SVM tasks.}
  \label{fig:SVM-task-freq}
\end{figure}

To assess participants’ familiarity with the SVM life cycle, we asked them to rate their involvement across four SVM tasks. As shown in Figure \ref{fig:SVM-task-freq}, all four tasks were well represented. SV detection emerged as the most common engagement, with 63\% of participants indicating they contributed “a lot” or “a great deal.” Assessment followed closely (55\%), while repair (50\%) and disclosure (53\%) showed slightly lower levels of active involvement. These results confirm that participants are not only broadly engaged with SVM tasks but that many operate across multiple stages of the workflow.
\vspace{-2.5mm}

\begin{figure}[h]
  \centering
  \includegraphics[width=0.95\linewidth]{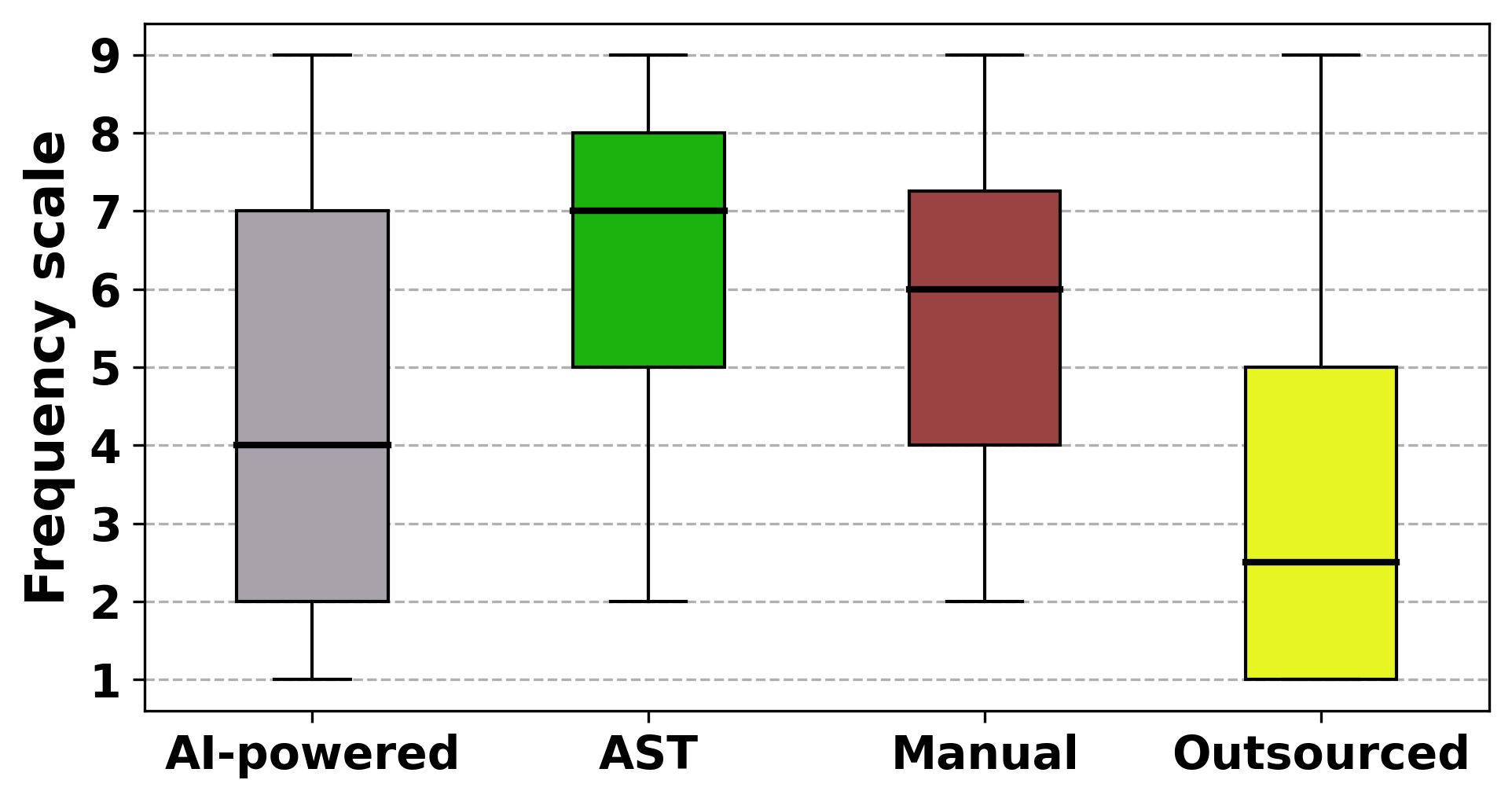}
  \caption{Frequency of use across four SVM approaches.}
  \label{fig:SVM-approaches}
\end{figure}

Figure \ref{fig:SVM-approaches} presents the frequency distribution of four common SVM approaches used by participants. AST tools were the most frequently adopted, with a median frequency of 7 ("most of the time") and a relatively narrow interquartile range (IQR), indicating consistent usage across respondents. Manual review was also common (median= 6), with a slightly wider IQR (4–7), suggesting more variation in how heavily different practitioners rely on it. In contrast, AI-powered tools showed a much wider spread in responses (IQR spanning 2 to 7), with a median of 4 ("many times"), reflecting more heterogeneous adoption. Outsourced services were the least utilized approach, with most participants using them only rarely (median= 2), suggesting that most organizations handle SVM in-house. Through this question, we also identified that 51 out of the 60 participants (85\%) reported some level of experience using AI-powered tools for SVM, while 9 participants selected “never,” indicating no such experience. Accordingly, the analyses and results related to AI usage presented in the remainder of the paper are based on the responses of the remaining 51 participants who reported at least some usage.
\vspace{-3mm}
\vspace{-2mm}

A Friedman test \cite{friedman1937use} applied to the nine-point usage scales indicated a significant difference across the four SVM approaches ($\chi^2(3) = 88.2$, $p < .001$). Post-hoc Wilcoxon signed-rank tests with Holm correction \cite{holm1979simple} showed a clear usage ranking: AST tools were used most frequently, followed by manual review, then AI-powered tools, and finally outsourced services. Each adjacent pairwise comparison was statistically significant at $p < .01$. Although participants reported regular use of AI-based tools, traditional approaches such as AST tools and manual code review remain more prevalent. This highlights that AI is not yet the primary approach for SVM, but rather complements existing approaches. We also calculated Kendall’s W \cite{kendall1939problem} ($W = 0.49$) to quantify the agreement in participant rankings. This coefficient ranges from 0 (no agreement) to 1 (complete agreement). A value of 0.49 indicates moderate consensus \cite{kendall1939problem}, meaning the observed usage pattern is reasonably consistent across respondents rather than being skewed by outliers.

\subsection{RQ1: Usage of AI-Powered Tools in SVM}
\label{sec:RQ1}

\subsubsection{\textbf{AI-Powered Tools in Use.}}
\label{sec:AI Tools in Use}
\begin{figure}[h]
  \centering
  
  \begin{minipage}[t]{0.48\linewidth}
    \centering
    \includegraphics[width=\linewidth]{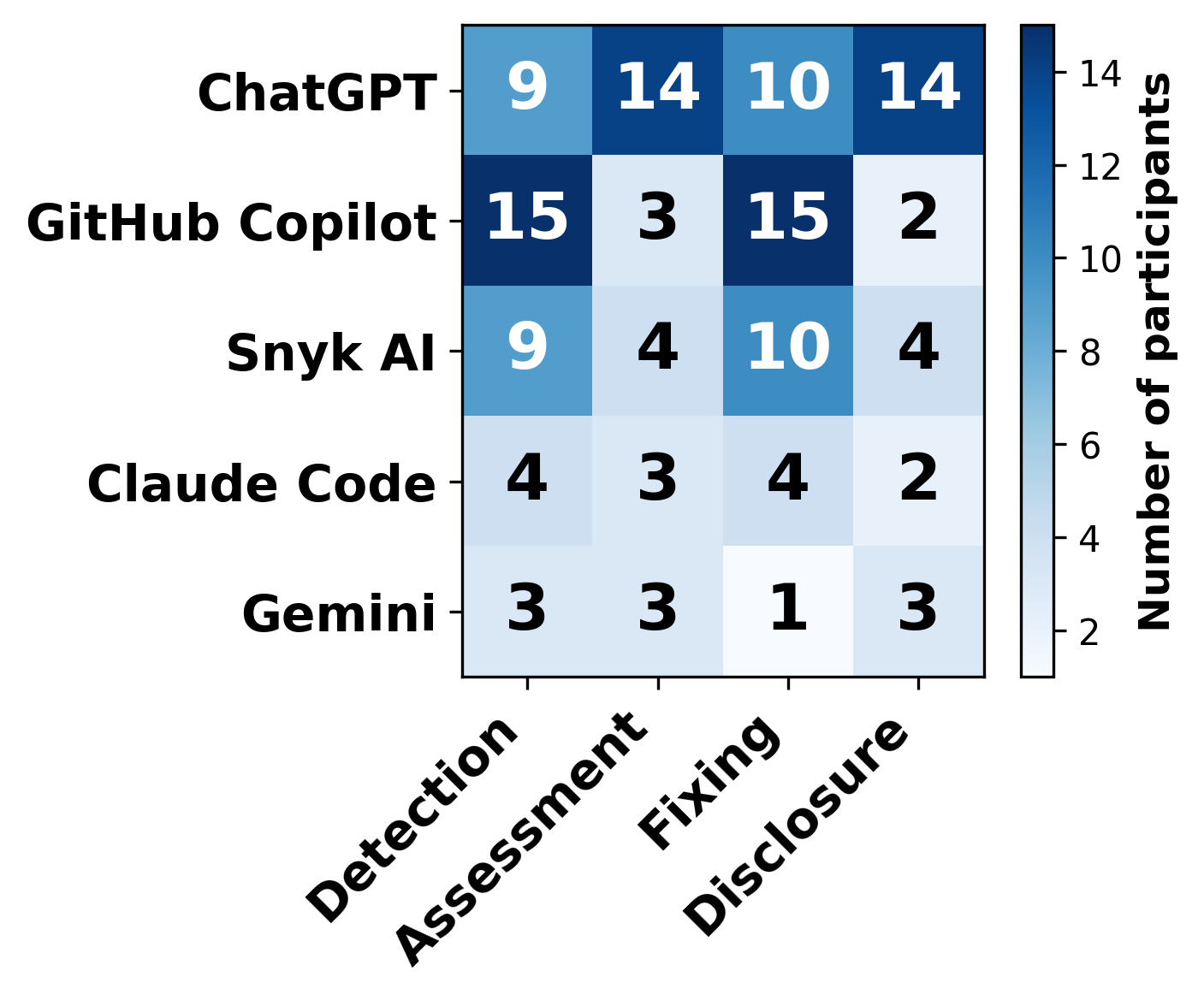}
    \caption{Top-5 AI tools used per SVM task (number of users).}
    \label{fig:ai-tools}
  \end{minipage}\hfill
  \begin{minipage}[t]{0.48\linewidth}
    \centering
    \includegraphics[width=\linewidth]{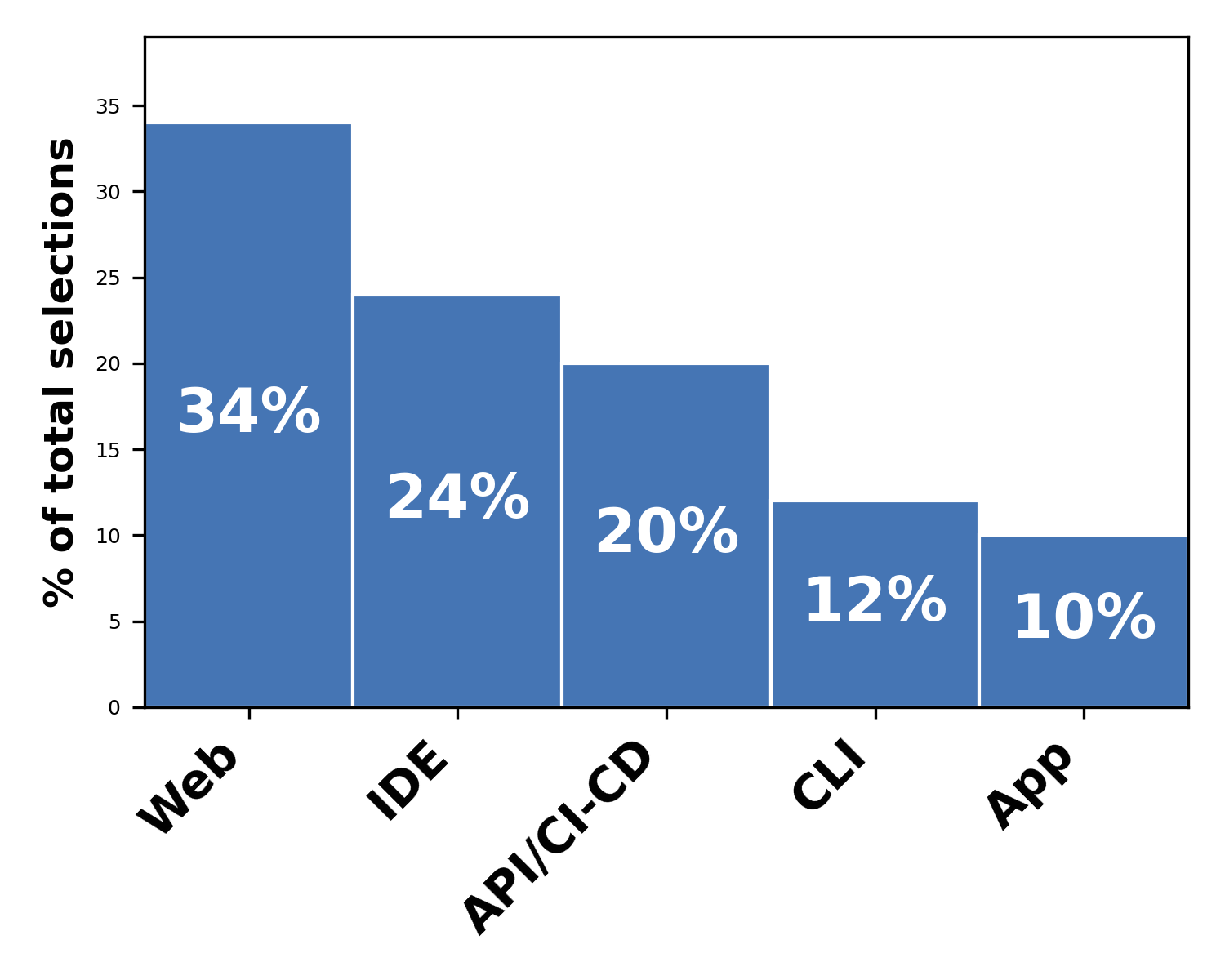}
    \caption{Distribution of tool access modes for AI-powered SVM (\% of total selections).}
    \label{fig:access-modes}
  \end{minipage}

\end{figure}

Participants listed the 
tools they use for each of the four main SVM tasks. As shown in Figure~\ref{fig:ai-tools}, ChatGPT \cite{ChatGPT} and GitHub Copilot \cite{GitHub-Copilot} were the most commonly mentioned tools. 
ChatGPT was frequently used for assessment and disclosure (14 participants each), reflecting its strength in explanation and communication tasks. GitHub Copilot led usage in detection and fixing (15 participants each), aligning with its integration in developer environments and ability to support code generation and patching. Snyk AI \cite{snykAI} was also notable for detection and fixing, consistent with its focus on automated security scanning and remediation. Other tools like Claude Code \cite{claudeCode} and Gemini \cite{google2023gemini} saw moderate use across tasks. Overall, the figure shows that AI tools are used across all core SVM tasks, confirming that adoption spans the full life cycle rather than a single phase. The complete list of tools is provided in the supplementary materials \cite{figshare}.
\vspace{-2mm}

\vspace{-3mm}
\subsubsection{\textbf{Tool Access Modes.}}
\label{sec:Tool Access Modes}
Participants were asked how they typically accessed AI-powered tools for SVM tasks. As shown in Figure \ref{fig:access-modes}, the most common access mode was through web-based interfaces (\color{black}34\color{black}\%), indicating a strong preference for ease of use and platform independence. Integration within IDEs (\color{black}24\color{black}\%) and CI/CD pipelines or custom systems via API (\color{black}20\color{black}\%) were also popular, reflecting the value of embedding AI capabilities directly into development workflows. Less frequently, participants interacted with AI tools via command-line interfaces (\color{black}12\color{black}\%) or standalone applications (\color{black}10\color{black}\%).

\subsubsection{\textbf{Influence on Decision-Making.}}
\label{sec:Influence_on_Decision_Making}
Participants were asked how AI-powered tools influence their decision-making in SVM tasks. The vast majority (\color{black}80\color{black}\%) indicated that these tools primarily served as a source of recommendations that still required human review and validation. A smaller portion (\color{black}14\color{black}\%) noted that AI tools mostly supplemented existing methods without directly affecting decisions. Only \color{black}6\color{black}\% reported using AI to automate decision-making with minimal human oversight. These findings suggest that most practitioners currently consider AI as a supportive aid rather than a replacement for expert judgment.

\subsubsection{\textbf{Workflow Integration Practices.}}
\label{sec:Workflow_Integration_Practices}
We asked participants to outline how they incorporate the output of AI-powered tools into their SVM workflow. 
A qualitative analysis of their open-ended responses yielded three themes, shedding light on the ways security teams embed AI assistance within their established practices.

\textbf{\bcircled{T1} Human-Centered Governance and Oversight} (\#Refs~=~42).
This theme captures the pervasive insistence that people, not AI models, remain the final authority. The dominant subtheme, \textbf{\emph{Human Validation and Oversight}} (\#Refs~=~42), shows practitioners routinely cross-checking AI findings, consulting official documentation, and reproducing exploits before actioning a fix \color{black}(SVM-specific verification step)\color{black}. As one participant put it, \emph{“Almost everything that is AI-generated must be scrupulously fact-checked”} (P2).

\textbf{\bcircled{T2} Risk-Driven and Collaborative Workflow Integration} (\#Refs = 39). This theme reveals how teams translate AI insights into action. The \textbf{\emph{Risk-Based Prioritization}} subtheme (\#Refs~=~12) shows practitioners ranking issues via CVSS scores, asset criticality, and business context \color{black} (SVM-specific triage criteria)\color{black}: \emph{“We triage and classify SVs by severity and affected systems, creating tickets with appropriate priority levels”} (P6). The companion subtheme, \textbf{\emph{Collaborative Remediation, Documentation, and Knowledge Sharing}} (\#Refs~=~27), highlights cross-functional coordination: \emph{“Once verified, I log it in Jira, loop in the right stakeholders, and tweak the remediation to match our coding standards”} (P32). These practices ensure scarce resources target the most consequential threats while building a reusable knowledge base for future incidents. 

\textbf{\bcircled{T3} Integrated Automation for Secure DevOps} (\#Refs~=~27). This theme describes a technical backbone where AI outputs flow into existing DevOps tooling. Participants embed AI-powered code scanners into CI/CD gates, SIEM dashboards, and automated pull-request workflows \color{black}(SVM-specific gating in delivery pipelines)\color{black}. One respondent explained, \emph{“We integrate AI findings into CI/CD pipelines for continuous security checks and use automated pull requests with suggested code fixes”} (P13). Others stressed the closed-loop nature of this pipeline: \emph{“After the tool generates the patch, I review the auto-PR, run tests in staging, then merge and re-scan”} (P46). Subthemes such as \textbf{\emph{Workflow Integration and Tracking Automation}} (\#Refs~=~12) and \textbf{\emph{AI-Assisted Testing and Auto-Remediation}} (\#Refs~=~9) together depict an ecosystem where detection, assessment, patch generation, and reporting occur with minimal manual hand-offs, yet still allow human override for high-risk exceptions. 

Collectively, these themes portray practitioners who balance automation with accountability: they welcome AI for speed and scale, but only when governance, risk prioritization, and collaborative safeguards are firmly in place.

\subsubsection{\textbf{Use and Validation of LLMs.}}
\label{sec:Use_and_Validation_of_LLMs}
\begin{figure}[htbp]
  \centering
  \includegraphics[width=0.98\linewidth]{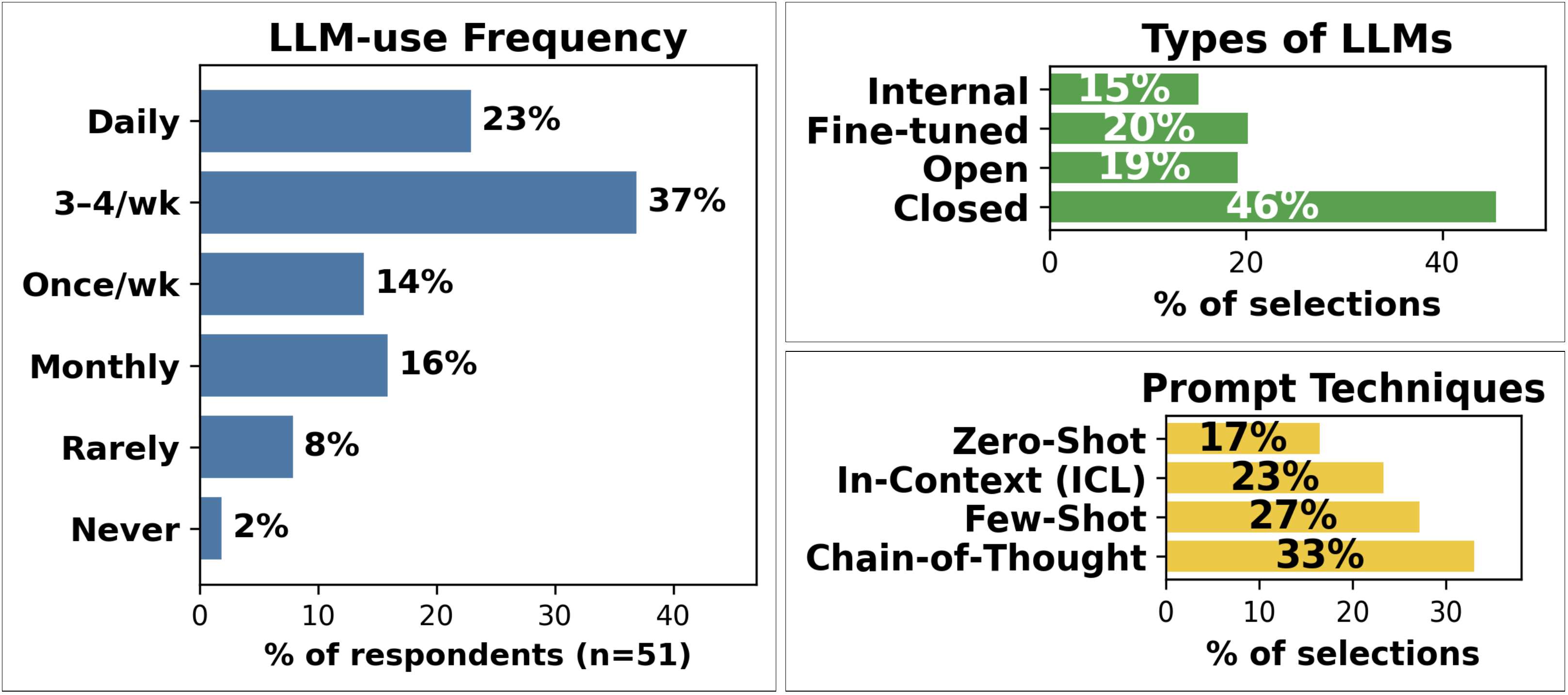}
  \caption{Usage patterns of LLMs in SVM tasks, including frequency of use, 
  model types, and prompting techniques.}
  \label{fig:LLM-3Pics}
\end{figure}

\color{black}This subsection focuses on LLM-specific usage and validation for SVM tasks. Workflow embedding that applies across AI tools is discussed in Section~\ref{sec:Workflow_Integration_Practices}. \color{black}
As noted in our survey design (Section~\ref{sec:AI Practices}), we included a dedicated set of questions on LLMs due to their frequent mention during the pilot phase. Participants reported how they use LLMs in SVM tasks, including usage frequency, model types, prompting techniques, and validation practices. As shown in Figure~\ref{fig:LLM-3Pics}, LLM usage was frequent across the sample: \color{black}60\color{black}\% of respondents used them at least 3–4 times per week, including \color{black}23\color{black}\% using them daily. This suggests that LLMs are a routine part of many participants’ workflows.

When asked about the types of LLMs they primarily use for SVM tasks, nearly half of respondents (\color{black}46\color{black}\%) reported using closed-source commercial LLMs such as ChatGPT or Gemini, suggesting a strong preference for readily accessible platforms. Fine-tuned LLMs (\color{black}20\color{black}\%) referred to models that had been further adapted with domain-specific data or organizational context, often built on top of existing base models. A similar proportion (\color{black}19\color{black}\%) reported using open-source models (e.g., LLaMA or Falcon) in their original forms. While open models offer greater transparency and customization potential, they require more technical overhead. Internally developed LLMs, those built and trained from scratch within the organization, were the least common (\color{black}15\color{black}\%), reflecting the significant resources and expertise needed for in-house model development. Prompting techniques varied: Chain-of-Thought prompting \color{black}(encourages the model to reason step-by-step ~\cite{ibm_chain_of_thought_prompting}) \color{black} was the most widely adopted (33\%), followed by Few-Shot \color{black}(provides a few input–output examples in the prompt ~\cite{ibm_few_shot_prompting}) \color{black} (\color{black}27\color{black}\%), In-Context Learning \color{black}(has the model learn from rich task-relevant context in the prompt without updating its parameters ~\cite{ibm_in_context_learning}) \color{black} (\color{black}23\color{black}\%), and Zero-Shot \color{black}(specifies the task in natural language with no examples ~\cite{ibm_zero_shot_prompting}) \color{black} (\color{black}17\color{black}\%). This diversity suggests that practitioners are experimenting with multiple techniques to extract more accurate or actionable outputs.

\vspace{-2.5mm}

\vspace{-2.5mm}
We also asked participants to describe the specific measures they took to verify the reliability of LLM outputs. Our qualitative analysis of their free-text answers surfaced three underlying themes, providing fresh insight into how software security practitioners build or withdraw trust in LLM recommendations \color{black}for SVM tasks\color{black}.

\textbf{\bcircled{T1} Human–Centric Verification and Governance} (\#Refs~=~56).
Participants begin from an attitude of measured skepticism and then rely on expert judgement, hands-on inspection and collegial oversight to legitimize or dismiss LLM advice.
Within this theme, the two most frequently cited subthemes were \textbf{\emph{Skepticism and Trust Calibration}} (\#Refs~=~14) and \textbf{\emph{Expert-Led Verification Workflows}} (\#Refs~=~29).
Practitioners openly voiced their reluctance to accept LLM output at face value—\emph{``I don't fully trust LLM outputs without additional measures''} (P1). Others underscored the primacy of personal expertise: \emph{``Just based on my feelings—if it makes sense to me, I trust it in some limited way''} (P19).
Fine-grained manual review was often augmented by peer checks—\emph{``Every output is reviewed by my team before it is acted on''} (P14)—showing how accountability shifts from an individual to a collective mode of governance.

\textbf{\bcircled{T2} Multi-Layer Technical Validation} (\#Refs~=~52).
Here practitioners triangulate evidence from standards, complementary tools, and empirical testing to confirm or refute an LLM’s findings \color{black}about suspected vulnerabilities\color{black}. Three prevalent subthemes structure this process: \textbf{\emph{Standards and Authoritative Source Validation}} (\#Refs~=~18) \color{black} (checking against security standards such as OWASP or NIST)\color{black}, \textbf{\emph{Triangulation with Tools and Models}} (\#Refs~=~17)    \color{black}(cross-checking with security tools like SonarQube)\color{black}, and \textbf{\emph{Isolated Testing and Vulnerability Reproduction}} (\#Refs~=~17) \color{black} (reproducing the vulnerability in a sandbox before proceeding)\color{black}. Standards alignment was routine—\emph{“I always verify LLM outputs against OWASP, NIST, etc.”} (P6)—while tool-level triangulation mitigated false positives—\emph{“Validate AI-detected vulnerabilities using manual inspection and tools like SonarQube or Snyk”} (P13). Empirical reproduction closed the loop: \emph{“If the LLM detects an exploit, we try to reproduce it in a sandbox; if we can’t, it’s a false positive”} (P17).

\textbf{\bcircled{T3} Continuous Risk Oversight and Constraints} (\#Refs = 9).
Ongoing assurance depends on adaptive safeguards but is also limited by real-world constraints.
The subtheme \textbf{\emph{Feedback Loops and Automated Safeguards}} (\#Refs~=~8) highlights iterative improvement—\emph{``When false positives or negatives are found, we retrain the model and add missed patterns to the fine-tuning dataset''} (P45) \color{black}to improve the accuracy of LLM outputs used across SVM tasks\color{black}—while \textbf{\emph{Resource and Access Constraints}} (\#Refs~=~1) reminds us that verification is not cost-free—\emph{``Access restrictions on certain tools or APIs can slow things down, often taking an hour or more''} (P7) \color{black}and can delay SVM verification timelines\color{black}.

These themes show that software security practitioners do not passively accept LLM outputs. Instead, they apply layered validation processes that combine human judgment, standards compliance, and evidence-based checks \color{black}to decide whether LLM results enter SVM workflows or are discarded\color{black}.

\subsection{RQ2: Perceived Strengths and Limitations of AI Tools in SVM}
\label{sec:RQ2}
\begin{figure}[h]
  \centering
  \includegraphics[width=0.98\linewidth]{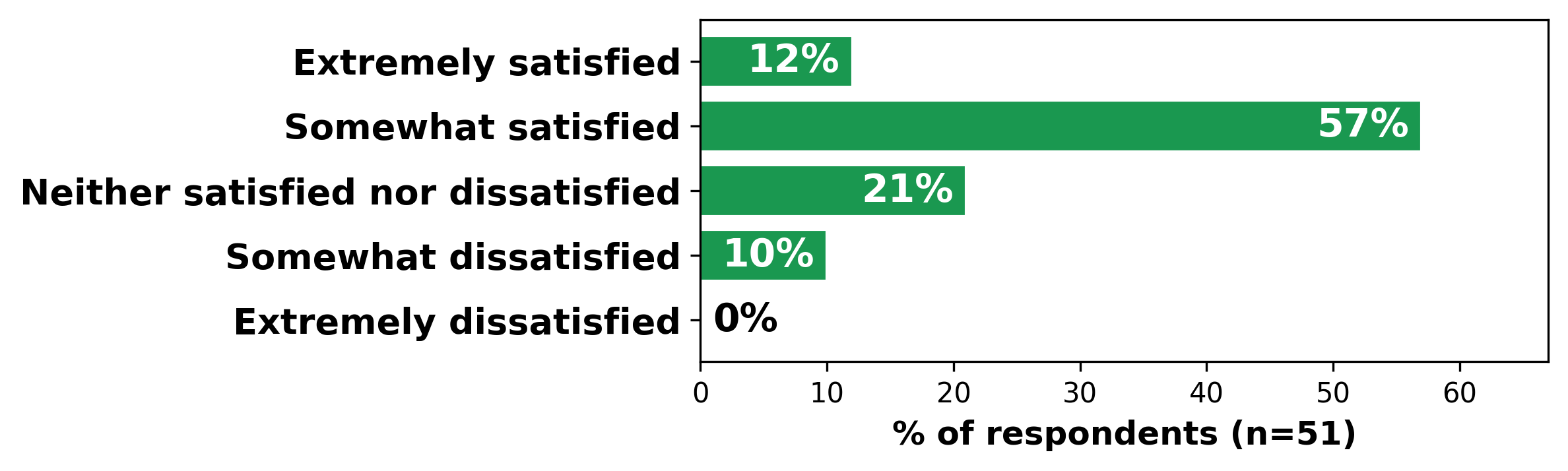}
  \caption{Overall satisfaction with AI-powered tools for SVM.}
  \label{fig:AI-satisfaction}
\end{figure}

We first asked participants to reflect on their overall experience using AI-powered tools for SVM. As shown in Figure~\ref{fig:AI-satisfaction}, the majority expressed a positive outlook: \color{black}57\color{black}\% reported being "Somewhat satisfied" and \color{black}12\color{black}\% "Extremely satisfied." Neutral views were also common (\color{black}21\color{black}\%), while dissatisfaction was relatively low, with only \color{black}10\color{black}\% indicating they were "Somewhat dissatisfied" and no participants selecting "Extremely dissatisfied."
\vspace{-3mm}
\vspace{-3mm}

These responses suggest that most practitioners find AI-powered tools beneficial, though there is still room for improvement. In the remainder of this section, we present the main strengths and limitations participants reported in open-ended responses.

\subsubsection{\textbf{Strengths.}}
\label{sec:Strengths}

We asked participants to describe, from their own experience, the key strengths of AI-powered tools for SVM. A qualitative analysis of their open–ended responses resulted in four themes, highlighting on how practitioners perceive the added value of AI across the SVM life cycle. 

\textbf{\bcircled{T1} Intelligent Detection and Threat Awareness} (\#Refs = 41).  
This theme describes how AI helps improve situational awareness by broadening detection coverage while 
reducing noise. It is anchored by the subthemes \textbf{\emph{Automated Vulnerability Scanning and Detection}} (\#Refs~=~17) \color{black} (SVM-specific) \color{black} and \textbf{\emph{Reduction of False Positives}} (\#Refs~=~9).  
Participants praised AI analytics for “\emph{filtering out 
noise and focusing on real threats}” (P1) and for surfacing issues that classical tools missed: \emph{``Copilot helped identify a subtle authentication-bypass vulnerability that traditional SAST had missed''} (P6).  Real-time behavioral engines further impressed respondents by catching zero-day activity: \emph{``AI-powered tools 
detect unknown vulnerabilities so they can be remediated before exploitation''} (P37).

\textbf{\bcircled{T2} Embedded Security Automation in the Development Life cycle} (\#Refs~=~38).  
Here practitioners emphasized the seamless integration of AI into CI/CD pipelines and IDEs, enabling security-as-code \color{black} (SVM gates in CI/CD and code review)\color{black}. Dominant subthemes include \textbf{\emph{In-Editor Secure Coding Assistance}} (\#Refs~=~16) \color{black} (vulnerability-aware suggestions) \color{black} and \textbf{\emph{Automated Remediation and Patch Application}} (\#Refs~=~14) \color{black} (SVM-specific)\color{black}.  For example, one engineer noted how, during code review, \emph{``the AI not only fixed the vulnerability I reported but detected others and patched them automatically, catching issues early in the pipeline''} (P5).  Another highlighted the time savings of automatic pull-request generation: \emph{``Snyk AI auto-generates pull requests, cutting patch time from nine hours to twelve minutes''} (P45).  IDE companions also proved valuable: \emph{``GitHub Copilot understands the feature I’m working on and suggests possible fixes or vulnerabilities''} (P17).

\textbf{\bcircled{T3} Contextual Decision Support and Knowledge Enablement} (\#Refs~=~27).  
Beyond raw detection, AI provides rich context that shapes remediation strategy. Key subthemes are \textbf{\emph{Risk-Based Prioritization}} (\#Refs~=~6) \color{black} (SVM-specific)\color{black}, \textbf{\emph{Simplified Reporting and Communication}} (\#Refs~=~12), and \textbf{\emph{Educational and Knowledge Support}} (\#Refs~=~9) \color{black} (explaining vulnerability issues)\color{black}.  Participants valued tools that rank SVs by business impact—\emph{``Rapid7 gives us a priority patch list instead of the whole CVE dump''} (P27)—and that translate security jargon: \emph{``ChatGPT explains issues in simple words, even to developers who aren’t security experts''} (P18).  One senior engineer summarized the benefit as \emph{``AI-powered fix suggestions have been a game-changer''} (P30).

\textbf{\bcircled{T4} Efficiency, Scalability and Productivity Gains} (\#Refs~=~22).  
Automation yields measurable performance improvements, as reflected in the subthemes \textbf{\emph{Workflow Acceleration}} (\#Refs = 11) \color{black} (faster vulnerability handling) \color{black} and \textbf{\emph{Scalability and Large Codebase Handling}} (\#Refs~=~11) \color{black} (scaling SVM across repositories)\color{black}.  Practitioners repeatedly cited drastic time savings: \emph{``Tasks finish 50\% earlier because understanding and anomaly hunting are automated''} (P25). Others stressed AI’s ability to tame massive repositories: \emph{``The AI chat helps me locate features in a large code base easily''} (P28).  Such efficiencies also improved engagement: \emph{``With fewer false flags, developers are finally willing to look into findings''} (P51).

Together, these themes depict an ecosystem in which AI is already augmenting practitioners’ capabilities—improving detection quality, embedding security into development workflows, and democratizing expertise \color{black} across the SVM life cycle\color{black}.

\subsubsection{\textbf{Limitations.}}
\label{sec:Limitations}
We 
asked participants to describe the 
challenges they encountered when using AI-powered tools for SVM. A qualitative analysis of their open-ended responses revealed three themes, offering insight into the technical, operational, and organizational frictions that 
constrain the adoption of AI in SVM.

\textbf{\bcircled{T1} Reliability and Accuracy of AI Vulnerability Management} (\#Refs = 91).
This theme captures doubts about the soundness of AI outputs.
The most frequently cited issue was \textbf{\emph{False positives, hallucinations and overstated risk}} (\#Refs = 31). One participant complained: \emph{``The main problem is that AI has given “ghost results” … it hallucinates and thinks something exists in the codebase when it really doesn’t''} (P47). Another spent half a day chasing a phantom CVE: \emph{``Snyk AI flagged CVE-2023-1234 … after five hours of manual testing I found out it was a non-issue''} (P45)\color{black}, delaying remediation decisions\color{black}.
Closely related is the problem of \textbf{\emph{Incomplete threat detection and coverage}} (\#Refs = 21). Several respondents stressed that models lag behind new or domain-specific exploits: \emph{``These tools miss vulnerabilities that are not in their training data … they’re not good for complex logic bugs''} (P38). Another added: \emph{``They struggle with zero-day and new threats''} (P1)\color{black}, creating SVM coverage gaps\color{black}.
A third prevalent subtheme, \textbf{\emph{Contextual blindness and misaligned recommendations}} (\#Refs = 23), reflects complaints that generic advice overlooks organizational realities: \emph{``The AI had trouble understanding the unique characteristics of our financial infrastructure … recommendations did not always align with regulatory requirements''} (P29)\color{black}, undermining SVM decisions in regulated environments\color{black}. 
Finally, \textbf{\emph{Quality and safety of AI-generated remediation}} (\#Refs = 16) shows how supposedly helpful fixes can break code: \emph{``Bolt suggested a patch which looked correct, but the fix had broken a separate piece of functionality and I spent extra time rolling it back''} (P5)\color{black}, reducing confidence in AI-generated SV patches\color{black}.

\textbf{\bcircled{T2} Operational and Resource Burden of AI Integration} (\#Refs = 34).
Despite promises of efficiency, AI often introduces new workload. The dominant subtheme is \textbf{\emph{Integration, performance and resource burdens}} (\#Refs = 22); practitioners reported costly pipelines, token limits and slow builds: \emph{``LLMs … have limitation in input- and output-tokens so you have to give them small chunks … sometimes they get into a loop because they forget what they just tried''} (P49). AI adoption therefore still demands substantial \textbf{\emph{Human oversight and effort}} (\#Refs = 12). Verification remains indispensable: \emph{``AI tools do help … but they can flag patterns as false positives … so double intervention is required''} (P15). These hidden costs temper the time-savings claimed by vendors.

\textbf{\bcircled{T3} Governance, Trust and Organizational Risk} (\#Refs = 14).
Beyond technical limits, respondents highlighted policy and trust barriers. Confidentiality restrictions, licensing costs and uncertainty in AI advice come together under the subtheme \textbf{\emph{Trust, compliance and organizational concerns}} (\#Refs = 14). For example, a bank security engineer noted: \emph{``Due to internal policies, LLMs hosted on public servers cannot be used … the solution would be an internal tool but that would generate additional cost''} (P3). Others voiced skepticism: \emph{``…unclear privacy policies… we can never fully depend on these tools for security''} (P18).

These themes reveal that while AI tools promise transformative support for SVM, practitioners still grapple with accuracy gaps, operational overhead, and governance hurdles that must be addressed before widespread, confident adoption is achievable. Some features were seen as both helpful and problematic depending on context, task, or team maturity. We conclude that these thematic tensions do not indicate contradictions, but rather reflect the evolving and transitional nature of AI integration in real-world SVM practice.

\section{Discussion}
\label{sec:Discussion}
In this section, we discuss our findings in relation to developer-centered SVM studies \cite{fu2024aibughunter,steenhoek2025closing,klemmer2024using,liu2024exploring} and \color{black} broader AI for SE perception studies beyond SVM \cite{vaithilingam2022expectation,ziegler2022productivity,sergeyuk2025using,liang2024large}, \color{black} and we outline key implications for practitioners, tool designers, and researchers.

\subsection{AI for SVM in Practice: How Far Are We?}
\label{sec:AIforSVM}
Our findings indicate that AI-powered tools are no longer confined to isolated PoCs solutions for SVM. Industry practitioners in our survey reported applying them at multiple stages of the SVM life cycle. 85\% of respondents had at least some experience with these tools, and their use covered detection, assessment, repair, and disclosure rather than a single task. This breadth complements Fu et al.'s evaluation of AIBugHunter ~\cite{fu2024aibughunter}, which demonstrated real-time IDE-integrated assistance for C/C++ SVs. However, the controlled setting of their study did not reveal several common operational challenges reported by our participants, including false positives, contextual blindness, and manual verification efforts. Such findings suggest that laboratory success metrics should be paired with deployment-oriented indicators such as human validation effort before practitioners can fully leverage the potential of AI.

Trust remains a decisive barrier between promising benchmarks and operational use. Both our findings and Steenhoek et al.'s study of DeepVulGuard ~\cite{steenhoek2025closing} identify false positives and lack of contextual awareness as significant barriers. Practitioners in our study consistently emphasized a layered validation approach involving manual inspection, sandbox testing, peer review, and cross-checking against established standards such as OWASP or NIST. Similarly, Klemmer et al. ~\cite{klemmer2024using} found that developers commonly inspect and test AI-generated code using multi-step validation processes. However, their participants were predominantly focused on general functional correctness rather than explicitly checking security aspects, raising concerns about potential complacency. Our results address these concerns by highlighting how governance checkpoints, such as peer-review gates and branch protection rules, are already embedded within organizational SVM workflows. This indicates that organizational safeguards, rather than individual caution, effectively calibrate trust in AI use within SVM.

Despite significant advances in applying AI for SVM, operational friction and equity concerns remain prominent.
Benchmark evaluations, such as Liu et al.'s ~\cite{liu2024exploring} study demonstrating ChatGPT’s superiority on controlled SVM tasks, contrast with our findings about real-world challenges. Participants reported integration issues such as prompt engineering complexity, token limits, licensing constraints, and infrastructure limitations. These challenges disproportionately affect smaller, resource-constrained organizations and risk widening existing disparities across the software security landscape. Whilst AI adoption for SVM has advanced within controlled environments, practical adoption remains constrained by operational realities and resource limitations.

In summary, AI-powered tools for SVM have made considerable progress toward broader industry adoption. However, challenges such as limited contextual awareness, socio-technical integration, and equitable access remain critical hurdles before these tools can reach their full potential in real-world settings.

\color{black}
\subsection{Comparison to Broader AI for SE}
\label{sec:relation-broader-ai-se}
In the broader AI for SE literature, developer perception studies predominantly examine \textit{code generation} (Section~\ref{Sec:AI4SE-beyondSVM}). These studies characterize AI assistants as providing helpful starting points rather than reliable accelerators, reporting no consistent time gains and persistent adoption barriers around trust, accuracy, and limited project context \cite{vaithilingam2022expectation,ziegler2022productivity,sergeyuk2025using,liang2024large}. 
In code generation, developers lack an established process for deciding whether to accept, modify, or reject AI code suggestions \cite{vaithilingam2022expectation,liang2024large}. Decisions range from quick acceptance to attempted repair or outright rejection, and occasional over-reliance is reported; these judgments are often constrained by developers’ difficulty in understanding AI-generated code \cite{vaithilingam2022expectation,liang2024large}.
By contrast, in SVM, the basis for action is assurance: as detailed earlier, AI outputs typically advance only after human verification, risk-based triage, and pipeline controls (Section~\ref{sec:Workflow_Integration_Practices}), with LLM-specific validation adding standards alignment, multi-tool corroboration, and sandbox testing (Section~\ref{sec:Use_and_Validation_of_LLMs}). As reported in Section~\ref{sec:Influence_on_Decision_Making}, most respondents (80\%) indicated that AI tools serve as recommendations that still require human review and validation, 14\% reported that AI mostly supplements existing methods, and only 6\% reported automating decisions with minimal human oversight. Consequently, SVM decisions are typically made only after required verification checkpoints, with 69\% of respondents reporting positive experiences in using AI-powered tools for SVM alongside these verification processes (Section~\ref{sec:RQ2}).
\color{black}

\subsection{Implications of the Findings}
\label{sec:Implications}

\underline{\textit{For Practitioners.}}
AI-powered tools can significantly enhance SVM effectiveness, provided they are embedded in workflows that preserve human oversight. Treating AI suggestions as probabilistic rather than definitive enables teams to safely harness automation without sacrificing reliability. Teams should enforce verification processes, adapt AI-generated outputs to internal standards, and integrate results into shared backlogs and review cycles. Leveraging organizational safeguards such as peer-review gates and branch protection rules can help mitigate potential individual complacency regarding AI-generated security suggestions.

\underline{\textit{For Tool Designers.}}
Tool adoption rarely follows a plug-and-play model. Participants reported various integration issues, including prompt tuning, token ceilings, licensing complications, and memory constraints. Designers should prioritize seamless integration with existing development pipelines and workflows. Key design opportunities include generating patches that align with project-specific coding styles, producing confidence-calibrated outputs, providing clear provenance links (e.g., CVE references), and incorporating feedback mechanisms that iteratively refine AI performance. Incorporating transparent explanations, audit logs, and governance-oriented features like confidence scores could further reduce mistrust and facilitate broader adoption.

\underline{\textit{For Researchers.}}
There is a clear need to move beyond benchmark-centric evaluations toward longitudinal field studies examining AI–human collaboration in realistic settings. Researchers should quantify not only patch accuracy but also socio-technical impacts such as developer effort, verification costs, and changes in team workflows. Future studies should investigate selective adoption and rejection patterns influenced by organizational constraints, including compliance policies, infrastructure limitations, and proprietary API access. Researchers should also explore cost-aware deployment strategies, lightweight models, local inference options, and fallback behaviors that support equitable adoption across organizations with different resource levels. Finally, researchers should examine design strategies that support bounded autonomy, such as incorporating human-in-the-loop approval for security-critical actions, so that AI complements rather than replaces human oversight.

\section{Threats to Validity}
\label{sec:Threats}

\textit{\textbf{External Validity.}} Our findings may not generalize to all software security professionals. Although our sample included 60 participants from 27 countries across six continents, it was drawn exclusively from a freelancing platform, which may skew toward professionals comfortable with freelance-based work. This could underrepresent those in highly regulated or enterprise settings. While the sample had broad geographic and role diversity, it may still overrepresent English-speaking professionals. Nonetheless, the variety of industries, roles, and seniority levels helps support the transferability of our findings to similar professional contexts.

\textit{\textbf{Internal Validity.}} We mitigated internal validity threats through several measures. To reduce misinterpretation of survey questions, we piloted and refined the survey with input from domain practitioners. For qualitative analysis, we collaboratively developed and refined a shared coding framework. During this process, we held regular discussions to ensure consistency across the dataset. We also monitored for thematic saturation, which was reached before the final responses. To ensure data quality, only participants with demonstrable SVM experience were retained.

\textit{\textbf{Construct Validity.}} Survey design followed established guidelines for empirical software engineering studies~\cite{linaaker2015guidelines} and was refined through expert piloting. However, the operationalization of constructs may not fully capture their complexity across diverse roles and contexts. For instance, participants’ interpretations of the four core SVM tasks may vary depending on their background or organizational setting. While we provided task descriptions to reduce ambiguity, some variation in understanding is likely. In addition, although some questions focused on perceptions (e.g., AI satisfaction level), we balanced these with questions on reported practices (e.g., AI output verification strategies) to provide a more comprehensive picture of AI engagement in SVM.

\section{Conclusions}
\label{sec:Conclusion}
In this study, we surveyed 60 experienced practitioners across 27 countries to examine how AI-powered tools are being used in SVM, their perceived strengths, and the challenges they present. Within our sample, 85\% of participants reported using such tools, highlighting their growing presence in the field. These tools are most commonly applied throughout the SVM life cycle, with strong engagement around LLM-based assistants. Practitioners highlighted benefits such as improved speed, broader coverage, and greater accessibility. However, these gains are tempered by concerns over trust calibration, contextual blind spots, operational frictions, and governance limitations.
When effectively integrated into risk-aware, human-governed workflows, AI-powered tools are seen as accelerators, enhancing expert judgment rather than replacing it.

Overall, our study offers a detailed view of how AI is currently supporting SVM efforts and provides actionable insights for practitioners, tool developers, and researchers working to ensure the safe and effective adoption of AI-powered tools in SVM.

\begin{acks}
This work has been supported by the Cyber Security Cooperative Research Centre Limited whose activities are partially funded by the Australian Government’s Cooperative Research Centre Program. We thank all participants for their time and contributions, and we acknowledge the Freelancer platform for supporting our recruitment efforts.
\end{acks}

\balance

\bibliographystyle{ACM-Reference-Format}
\bibliography{reference}

\end{document}